\documentstyle[prc,preprint,tighten,aps]{revtex}
\begin{document}
\draft
\preprint{\vbox{\noindent 
 \hfill LA-UR-97-131\\
 \null\hfill nucl-th/9701028}}
\title{Three-alpha structures in $\bbox{^{12}}$C}
\author{R. Pichler$^a$, H. Oberhummer$^a$, Attila
Cs\'ot\'o$^b$, S.~A. Moszkowski$^c$}
\address{$^a$Institut f\"ur Kernphysik, Technische 
Universit\"at Wien, Wiedner Hauptstra{\ss}e 8 - 10, 1040 
Vienna, Austria \\
$^b$Theoretical Division, Los Alamos National 
Laboratory, Los Alamos, NM 87545, USA \\
$^c$Physics Department, University of California at Los
Angeles, Los Angeles, CA 90024, USA}
\date{January 16, 1997}

\maketitle

\begin{abstract}
We search for three-alpha resonances in $^{12}$C by using
the complex scaling method in a microscopic cluster model.
All experimentally known low-lying natural-parity states 
are localized. For the first time we unambiguously
show that the $0^+_2$ state in $^{12}$C, which plays an 
important role in stellar nucleosynthesis, is a genuine 
three-alpha resonance.
\end{abstract}
\pacs{{\em PACS}: 21.45.+v, 21.60.Gx, 27.20.+n \\  
{\em Keywords}: $^{12}$C states, three-body resonances, 
few-body dynamics, three-alpha model, cluster model}

\narrowtext 

\section{Introduction}

Carbon, which is the fundamental basis of the chemistry of 
terrestrial life,  
is produced in red giant stars by burning the helium ash of 
hydrogen fusion. In order to produce carbon in stellar 
nucleosynthesis, the $A=5$ and $A=8$ nuclear mass-stability 
gaps must be bridged. Salpeter and \"Opik pointed out 
\cite{Salpeter} that the lifetime of $^8$Be is long enough, 
so that the $\alpha +\alpha \rightleftharpoons\; $$^8$Be 
reaction can produce macroscopic amounts of equilibrium 
$^8$Be in stars. Then, the unstable $^8$Be can capture an 
additional $\alpha$ particle to produce stable $^{12}$C. 
However, this so-called triple-alpha reaction has very 
low rate because of the low density of $^8$Be. 

Hoyle argued \cite{Hoyle} that in order to explain the 
measured abundance of carbon in the Universe, this 
reaction must proceed through a hypothetical resonance of 
$^{12}$C, thus strongly enhancing the cross section. 
Hoyle suggested that this resonance is a $J^\pi=0^+$ state 
at $E_{\rm r}=0.4$ MeV (throughout this paper $E_{\rm r}$ denotes 
resonance energy in the center-of-mass frame relative to 
the three-alpha threshold, while $\Gamma$ denotes the full 
width). Subsequent experiments indeed found a $0^+$ 
resonance in $^{12}$C in the predicted energy region. It 
is the second $0^+$ state ($0^+_2$) and the second 
excited state  of $^{12}$C. Its modern parameters 
$E_{\rm r}=0.3796$ MeV and $\Gamma=8.5\times10^{-6}$ MeV 
\cite{Ajzenberg} agree well with the old theoretical prediction.

We mention here that the long lifetime of the $^8$Be ground 
state and the existence of the $0^+_2$ resonance in $^{12}$C 
at the right energy region are only parts of an incredible 
chain of fortunate nuclear coincidences, which makes the 
abundant existence of carbon and oxygen possible. For an 
interesting account, see Ref.\ \cite{Rolfs}.

The aim of the present work is to explore the nature of the 
$0^+_2$ state in $^{12}$C. The facts that this state is 
very close to the three-alpha threshold, and that the alpha 
particle is strongly bound make it probable that the wave 
function of $0^+_2$ has a dominant three-alpha clustering 
nature. The low-lying states of $^{12}$C, including $0^+_2$, 
have been studied in a number of macroscopic (with 
structureless alpha particles) \cite{mac,Fedorov} and 
microscopic \cite{mic} three-alpha models. These models 
reproduce the general features of the low-lying $^{12}$C 
spectrum. However, all these models without exception 
assume three-body bound state- or two-body $^8{\rm Be}+
\alpha$ scattering state asymptotics for the wave 
functions. Thus, none of them obeys the physically correct 
three-body boundary condition for states above the 
three-alpha threshold. Those models which use $^8{\rm Be}+
\alpha$ asymptotics with bound-state-like $^8$Be \cite{mic} 
are seemingly adequate, because the small width of the 
$^8$Be ground state makes its wave function very similar 
to a bound state wave function in a large spatial region. 
However, one must realize that such a model predicts the 
states above the three-alpha threshold, e.g.\ $0^+_2$, 
to be two-body, $^8{\rm Be}+\alpha$, resonances. This means 
that currently there is no unambiguous evidence that 
these resonances are intrinsic states of $^{12}$C. In fact, 
it was speculated that the $0^+_2$ state is not a 
three-body resonance, but an enhancement coming from the 
$^{12}{\rm C}\rightarrow\:$$^8{\rm Be}+\alpha \rightarrow 
\alpha +\alpha +\alpha$ sequential decay \cite{reply}. 
This idea was, however, criticized \cite{comment} by 
arguing that all the experimental data supported the 
genuine $^{12}$C nature of this state. 

In the present paper we use a method which is able to 
handle the three-body dynamics of the $0^+_2$ state 
correctly. Thus for the first time we can unambiguously 
show whether this state is a genuine three-alpha resonance 
in $^{12}$C. We also study other low-lying natural-parity 
states of $^{12}$C.

\section{Model}

Our model is a microscopic three-cluster $(\alpha+ 
\alpha+\alpha)$ resonating group method (RGM) approach 
to the twelve-nucleon system. The trial function of the 
twelve-body problem has the form
\begin{equation}
\Psi^{^{12}{\rm C}}=\sum_{l_1,l_2} {\cal A} \Bigl 
\{ \Phi^\alpha \Phi^\alpha \Phi^\alpha\chi^{\alpha(
\alpha\alpha)}_{[l_1l_2]L} (\bbox{\rho}_1,\bbox{
\rho}_2) \Bigl\},
\label{wfn}
\end{equation}
where ${\cal A}$ is the intercluster antisymmetrizer, 
the $\Phi^\alpha$ cluster internal states are 
translationally invariant $0s$ harmonic-oscillator 
shell-model states with zero total spin, the $\bbox{\rho}$ 
vectors are the intercluster Jacobi coordinates, $l_1$ and 
$l_2$ are the angular momenta of the two relative motions, 
$L$ is the total orbital angular momentum and $[\ldots]$ 
denotes angular momentum coupling. The total spin and 
parity of $^{12}$C are $J=L$ and $\pi=(-1)^{l_1+l_2}$, 
respectively. 

Putting (\ref{wfn}) into the twelve-nucleon Schr\"odinger 
equation which contains the nucleon-nucleon ($N-N$) strong 
and Coulomb interactions, we get an equation for the 
intercluster relative motion functions $\chi$. These 
functions represent the three-body dynamics of the $^{12}$C 
states. In order to determine these functions, we have to 
use a method which can handle three-body resonances. 

There are indirect and direct approaches. The indirect 
methods, e.g.\ \cite{Danilin}, study the three-body problem 
at real energies, and extract resonance parameters from the 
three-body phase shifts. A similar method has recently been 
used to study the $0^+_2$ state of $^{12}$C \cite{Fedorov}. 
However, the resonant nature of this state and the 
resonance parameters were extracted from a bound state wave 
function by using the WKB approximation. 

The aim of the direct methods for resonances is to find 
the complex-energy poles of the three-body scattering 
matrix. For example, in \cite{Matsui,Eskandarian,Emelyanov} 
the authors determined the pole positions of the three-body 
$S$-matrix by analytically continuing the homogeneous 
Faddeev-equation to complex energies. To get a decisive 
answer to the question of the nature of the $0^+_2$ state, 
we must use a direct method. 

Our choice is the complex 
scaling method (CSM) \cite{CSM}. It reduces the problem 
of asymp\-to\-ti\-cal\-ly divergent resonant states to that of 
bound states, and can handle the Coulomb interaction 
without any problem. The main point of the CSM is, that 
instead of solving the original Schr\"odinger equation  
for resonances, a new Hamiltonian is defined by
\begin{equation}
\widehat{H}_\theta=\widehat{U}(\theta)\widehat{H}
\widehat{U}^{-1}(\theta),
\end{equation}
and the complex equation
\begin{equation}
\widehat{H}_\theta|\Psi_\theta\rangle=\varepsilon|\Psi_
\theta\rangle
\label{csh}
\end{equation}
is solved. In coordinate space the unbounded similarity 
transformation $\widehat{U}(\theta)$ acts on a function 
$f(r,\hat r)$ as
\begin{equation}
\widehat{U}(\theta)f(r,\hat r)=e^{3 i \theta /2}f(re^{i
\theta},\hat r),
\label{cs}
\end{equation}
where $\hat r$ describes the angular part of ${\bf r}$. For real 
angles $\theta$, $\widehat{U}(\theta)$ results in a 
rotation into the complex coordinate plane, whereas for 
complex $\theta$, $\widehat{U}(\theta)$ results in a 
rotation and scaling. Further on, we shall always use real 
$\theta$ values. In the case of a many-body problem, the 
transformation given by Eq.\ (\ref{cs})  has to be performed in each 
dynamical Jacobi coordinate.
 
For a broad class of potentials there is a remarkable 
connection between the spectra of $\widehat{H}$ and 
$\widehat{H}_\theta$ \cite{abc}: (i) the bound eigenstates 
of $\widehat{H}$ are eigenstates of $\widehat{H}_\theta$, 
for any value of $\theta$ within $0\leq \theta \leq \pi/2$;
(ii) the continuous spectrum of $\widehat{H}$ is  
rotated by an angle $2\theta$; (iii) the complex 
generalized eigenvalues of $\widehat{H}_\theta$, 
$\varepsilon_{\text{res}}=E_{\rm r}-i\Gamma/2$
(with $E_{\rm r},\, \Gamma >0$), belong to its 
proper spectrum, with square-integrable eigenfunctions, 
provided $2\theta>|{\rm arg}\, \varepsilon_{\text{res}}|$. 
These complex eigenvalues coincide with the $S$-matrix 
pole positions. 

In nuclear physics the CSM has been successfully applied 
to two-body problems, like in a RGM description of 
$^8$Be \cite{Be8}, in an OCM model of $^{20}$Ne 
\cite{KruppaKato} and in the OCM description of the 
resonances of $^{10}$Li \cite{IkedaKato}. It was also 
tested for three-body resonances \cite{tbr}, and was 
used, e.g., in a RGM model of $^6{\rm He}$, $^6{\rm Li}$ 
and $^6{\rm Be}$ \cite{soft}, in searching for 
three-nucleon resonances \cite{3n}, in three-body 
models of $^6$He, $^{10}$He and $^{11}$Li \cite{Kato} and 
in an RGM model of $^9$Be and $^9$B \cite{Arai}.  
Further details and references of the method can be found 
there. We note, that the CSM is identical to a contour 
rotation in momentum space \cite{Afnan}. The latter method 
was also used, for example, to study three-body resonances 
in the ${\rm A}=6$ nuclei \cite{Eskandarian}.
 
Up to Eq.\ (\ref{csh}) our treatment of three-body 
resonances is exact. Since the resonant wave functions 
become square-integrable in the CSM, we can use any 
bound-state method to describe them. We expand the relative 
motion functions $\chi$ in Eq.\ (\ref{wfn}) in terms of 
products of tempered Gaussian functions, $\rho_1^{l_1}
\exp [-(\rho_1/\gamma_i)^2]Y_{l_1m_1}(\widehat \rho_1)
\cdot \rho_2^{l_2} \exp [-(\rho_2/\gamma_j)^2]Y_{l_2m_2}
(\widehat \rho_2)$ (where $l_1$ and $l_2$ are the angular 
momenta in the two relative motions, respectively, and 
the widths $\gamma$ of the Gaussians are the parameters 
of the expansion), and determine the expansion coefficients 
from the $\langle\delta \Psi_\theta|\widehat{H}_\theta-
\varepsilon| \Psi_\theta\rangle=0$ projection equation. 
This way we discretize the continuum and select the 
square-integrable solutions of Eq.\ (\ref{csh}). We use 
ten Gaussian basis functions in each relative motion.  
The matrix elements of the complex scaled many-body 
Hamiltonians were calculated in exact analytic forms by 
using computer algebraic techniques. 

\section{Results}
 
In order to avoid any possible model dependence of the 
conclusions we use three different effective $N-N$ 
interactions. The Minnesota (MN) force was designed to 
reproduce low-energy $N-N$ scattering data \cite{MN}, 
while the rather different Volkov 1 (V1) and 2 (V2) forces 
were obtained from fitting the bulk properties of $s$- and 
$p$-shell nuclei \cite{Volkov}. Each force contains an 
exchange mixture parameter, $u$ and $m$, 
respectively. We fix these parameters by requiring 
that the energy of the $^8$Be ground state be reproduced. 
The harmonic-oscillator size parameters of the alpha 
particle internal states are chosen to minimize the 
free-alpha energies. Thus, the wave function of the alpha 
particle is variationally stabilized. The size parameters 
of the alpha particle and the exchange mixture parameters 
of the $N-N$ forces are listed in Table 1 for the three 
interactions, together with the energies and radii of the 
alpha particle.

The parameters of the low-lying $^8$Be resonances, given 
by the MN, V1 and V2 forces, are shown in Table 2. For 
$^8$Be we use a two-alpha cluster model wave function, 
similar to Eq.\ (\ref{wfn})
\begin{equation}
\Psi^{^8{\rm Be}}={\cal A} \Bigl \{ \Phi^\alpha
\Phi^\alpha \chi^{\alpha\alpha}_L(\bbox{\rho}) \Bigl\}.
\label{wfn8}
\end{equation}
The relative motion function $\chi$ is determined by 
using a two-body scattering approach based on the 
Kohn-Hulth\'en variational method \cite{Kamimura}. The 
$\alpha+\alpha$ scattering phase shifts, coming from the 
MN interaction, are shown in Fig.\ 1, together with the 
experimental data. A nice agreement is observed. The 
resulting scattering matrices are continued to complex 
energies, where their poles are localized \cite{pole}. 
The resonance parameters in Table 2 were extracted from 
the complex pole positions. 

This analytic continuation 
method is in principle equivalent with the CSM, but is 
numerically far more precise. The CSM has difficulties 
in localizing very narrow resonances. As an illustrative 
example, we performed CSM calculations for the $^8$Be 
resonances and found that although the energy of the 
ground state is well reproduced, the width is strongly 
overestimated. For $\theta$ values which give stable 
complex energy spectra, with the discretized continuum 
rotated by close to $2\theta$, this overestimation is 
more than two orders of magnitude. We could fine-tune 
the $\theta$ angle and the Gaussian basis parameters to 
get the experimental widths, but 
then the discretized continuum points would be scattered, 
forming a band rather than a line. If we used such a 
$\theta$ value in the $^{12}$C calculations, the results 
would be disastrous, making the identification of the 
resonances impossible. Therefore, we use such $\theta$ 
values in the $^{12}$C calculations which give more 
or less stable discretized continua. The price we 
have to pay for this is that we cannot resolve very 
small widths. However, this is not a serious problem, 
because our primary goal is to show the {\em existence} 
of the $^{12}$C states, and not to determine the precise 
resonance parameters. Our model and effective 
interactions are probably inadequate for this latter 
purpose.

The $N-N$ interactions, thus set to reproduce the unbound 
$^8$Be ground state, are used in the $^{12}$C calculations. 
In this way we ensure that there is no 
bound two-body subsystem in the three-alpha system as shown
also by the experimental data. Therefore,  our 
model handles the three-body dynamics properly, and the 
resonances we find are genuine three-alpha structures.

We perform calculations for the low-lying natural-parity 
states of $^{12}$C. In the three-alpha wave function 
(\ref{wfn}) we include $l_1=0$ for the $0^+$ and $2^+$ 
states, and additionally $l_1=2$ for the $1^-$ and $3^-$ 
states. Our test calculations show that the addition of 
further configurations hardly influences the results. 
For instance, if we include the $[l_1,l_2]L=[2,2]0$ 
configuration in the $0^+_1$ ground state wave function 
in addition to the $[0,0]0$ one, we gain less than 1\% 
in the three-body binding energy and less than 0.1\% in 
the absolute energy of $^{12}$C. This is in sharp contrast 
to the findings of macroscopic models \cite{mac} which 
assume structureless alpha particles. We believe that the 
antisymmetrization causes the $[0,0]0$ state to strongly 
dominate.

The parameters of the three-body resonances we find, using 
the MN, V1 and V2 forces are listed in Table 3, together 
with the experimental values. The ground state of $^{12}$C 
is strongly overbound by MN, roughly reproduced by V1 and 
underbound by V2. Interactions V1 and V2 give almost the 
same {\em absolute} energy for the ground state, so the 
difference between the energies relative to the three-alpha 
threshold comes from the different alpha particle energies 
shown in Table 1. However, this is not the case for the 
differences between the MN and V1 and V2 ground state 
energies. The reason of the differences between the MN 
and Volkov results is most probably the different exchange 
mixture structure of these forces. While the MN force 
reproduces the deuteron binding energy in an $L=0$ space, 
assumed here, the Volkov forces underbind the deuteron. 
However, the Volkov forces unphysically bind the singlet 
dinucleon states. 

We believe that the $^{12}$C ground state should be 
expected to be overbound in three-alpha models. The reason 
is that the ground state of $^8$Be is not a perfect 
two-alpha state. It has been shown that the inclusion of 
$^7{\rm Li}+p$ and $^7{\rm Be}+n$ channels in the ground 
state wave function of $^8$Be significantly increases its 
``binding energy'' \cite{Li7p}. Thus, in order to reproduce 
the $^8$Be ground state energy in a model which contains 
these configurations, the $N-N$ interaction (or the 
$\alpha-\alpha$ interaction in macroscopic models)  
should be weakened. We expect that then the $0^+_1$ state 
of $^{12}$C would be closer to the experimental position, 
using the MN force, even if the $^{12}$C wave function also 
contains higher-lying rearrangement channels. We note, 
that all macroscopic models underbind the $0^+_1$ state 
of $^{12}$C \cite{mac}. The only exception is Ref.\ 
\cite{Marsh} where the ``microscopic'' $\alpha-\alpha$ 
potential was constructed in a way which took into account 
some of the effects of the internal structure of the alpha 
particle. In agreement with our MN result, Ref.\ 
\cite{Marsh} found the ground state of $^{12}$C to be 
overbound.      

Table 3 shows that all experimentally known 
low-lying natural-parity states of $^{12}$C are reproduced 
by our model. This means that these states are all genuine 
three-alpha resonances. In many cases the calculated 
resonance parameters are far from the experimental values.
In order to get closer to the experiments, major
improvements of the model, e.g., the inclusion of
rearrangement channels with nonzero spin, would be
necessary. 

Our most important result is that 
we can localize the $0^+_2$ state as a three-body 
resonance. In Fig.\ 2 we show the low-energy part of the 
$0^+$ spectrum of the complex scaled Hamiltonian with the 
MN interaction. The two three-body resonances are denoted 
by the circles. The width of the $0^+_2$ state is probably 
overestimated by the CSM (cf.\ the discussion about the CSM 
and very small widths). The dots in Fig.\ 2 represent the 
rotated discretized three-body cut. Due to numerical 
precision problems these points form a band rather than a 
line. We attempted to optimize the Gaussian basis, but could not 
get a result cleaner than the one shown in Fig.\ 2. The numerical 
stability of the continuum points is rather sensitive to 
the level of precision maintained during the calculations 
of the complex many-body matrix elements. Although every 
matrix element was calculated from (very involved) 
algebraic expressions, no special attention was paid to 
the full optimization of the computational algorithm 
against the loss of numerical precision. Nevertheless, 
the identification of the resonances is unambiguous.  

We found that the parameters of the resonances are 
approximately independent of the rotation angle $\theta$ 
within a reasonable interval. We performed most of the 
calculations in Table 3 with $\theta=0.2$ rad, and checked 
several of them with $\theta=0.1$ rad. Figure 2 shows the 
result of a calculation with $\theta=0.1$ rad.  We have 
encountered some problems regarding the $3^-$ states. 
Using $\theta=0.2$ rad, one of these resonances can be 
lost. That is why we used $\theta=0.1$ rad for this 
state in Table 3.

In contrast to the ground state of $^{12}$C, the $0^+_2$ 
state is predicted by all three interactions at roughly 
the same position, and close to the experiment. This is 
not surprising, because the $0^+_2$ state, being so 
close to the three-alpha threshold, is expected to be a 
more perfect three-alpha state than the ground state.

We would like to note, that although the $0^+_2$ state is a
three-body resonance, it does not decay into an uncorrelated
three-alpha final state. Experiments show that the decay
proceeds predominantly through the $^8{\rm Be}+\alpha
\rightarrow \alpha+\alpha+\alpha$ sequential process
\cite{Freer}. The experimental upper limit for the 
contribution of the three-alpha decay to the alpha decay 
width of the $0^+_2$ state is less then 4\%. The dominance 
of the $^8{\rm Be}+\alpha$ decay over the $3\alpha$ one is 
the result of the difference between the relative 
phase spaces \cite{Freer}.

\section{Conclusion}

In summary, we have studied the resonances of $^{12}$C in a
microscopic three-alpha model. We used the complex scaling
method, which allowed us to describe the three-body Coulomb
dynamics correctly for resonances. We used three different
effective nucleon-nucleon interactions and their results
are consistent with each other. We have localized all
experimentally known low-lying natural-parity states in
$^{12}$C, although a better agreement with experiment would
require major improvements of our model. For the first time
we were able to unambiguously show that the $0^+_2$ state
of $^{12}$C, which plays an important role in the
astrophysical triple-alpha process, is a genuine
three-alpha resonance.

\acknowledgments

We want to thank the Fonds zur F\"orderung 
wissenschaftlicher Forschung in \"Osterreich (project 
P10361--PHY) for their support. The work of A.\ C.\ was 
performed under the auspices of the U.S. Department of 
Energy, and was also supported by OTKA Grant F019701. 
Early stages of this work were done at the National 
Superconducting Cyclotron Laboratory at Michigan State 
University, supported by NSF Grants PHY92-53505 and 
PHY94-03666. The computer algebraic calculations were 
performed at the Department of Theoretical Nuclear 
Physics of Universit\'e Libre de Bruxelles. The kind 
help of Prof.\ Daniel Baye is appreciated.

\begin{figure}
\caption{Calculated $\alpha+\alpha$ scattering phase 
shifts in the center-of-mass frame using the MN 
interaction. Experimental data are taken from Ref.\ 
\protect\cite{Ajzenberg}.}
\label{Be8ps}
\end{figure}

\begin{figure}
\caption{Low-energy eigenvalues of the complex scaled 
Hamiltonian of the $0^+$ three-alpha states in $^{12}$C. 
The dots are the points of the rotated discretized 
continuum, while the circles are three-alpha resonances. 
The rotation angle is 0.1 rad.}
\label{energy}
\end{figure}

\narrowtext
\begin{table}
\caption{Harmonic-oscillator size parameter $\beta$
of the $\alpha$ internal state, exchange mixture 
parameter $u$ or $m$ of the $N-N$ interaction, and 
the energy $E_\alpha$ and point nucleon rms radius of the free 
$\alpha$ particle.}
\label{params}
\begin{tabular}{lr@{}lr@{}lr@{}lr@{}l}
&\multicolumn{2}{c}{$\beta$ (fm$^{-2}$)} & 
 \multicolumn{2}{c}{$u$ or $m$} & \multicolumn{2}{c}
 {$E_\alpha$} (MeV)& \multicolumn{2}{c}{$r_\alpha$ (fm)} \\
\hline
MN & 0.&6060 & 0.&93344& --24.&687& 1.&36 \\
V1 & 0.&5291 & 0.&57286& --27.&085& 1.&46 \\
V2 & 0.&5284 & 0.&60126& --27.&957& 1.&46 \\
Experiment & \multicolumn{2}{c}{---} & \multicolumn{2}{c}
 {---} & --28.&269 & 1.&48 
\end{tabular}
\end{table}

\widetext
\begin{table}
\caption{Energies (relative to the two-alpha threshold)
and full widths of low-lying resonances in $^8$Be. All 
quantities are given in MeV.}
\label{be8energ}
\begin{tabular}{lr@{}lr@{}lr@{}lr@{}lr@{}lr@{}lr@{}lr@{}l}
&\multicolumn{4}{c}{MN}&\multicolumn{4}{c}{V1}&\multicolumn
 {4}{c}{V2} &\multicolumn{4}{c}{Experiment 
\protect\cite{Ajzenberg}} \\
\cline{2-5}\cline{6-9}\cline{10-13}\cline{14-17}
&\multicolumn{2}{c}{E} & \multicolumn{2}{c}{$\Gamma$} &  
 \multicolumn{2}{c}{E} & \multicolumn{2}{c}{$\Gamma$} &
 \multicolumn{2}{c}{E} & \multicolumn{2}{c}{$\Gamma$} &
 \multicolumn{2}{c}{E} & \multicolumn{2}{c}{$\Gamma$} \\
\hline
$0^+$ & 0.&092 & \multicolumn{2}{c}{6.15$\times$10$^{-6}$} & 
 0.&092 & \multicolumn{2}{c}{2.36$\times$10$^{-6}$} & 
 0.&092 & \multicolumn{2}{c}{5.17$\times$10$^{-6}$} & 
 \multicolumn{2}{c}{0.09189} & \multicolumn{2}{c}{
 (6.8$\pm$1.7)$\times$10$^{-6}$} \\
$2^+$ & 3.&03 & \ \ 1.&39 & 2.&34 & \ \ 1.&48 & 2.&26 & 
\ \ 1.&42 &
\multicolumn{2}{c}{3.04$\pm$0.03} & \multicolumn{2}{c}{
1.50$\pm$0.02} \\
$4^+$ & 13.&10 & \ \ 4.&11 & 9.&96 & \ \ 5.&89 & 9.&55 &
\ \  5.&93 & 
\multicolumn{2}{c}{11.4$\pm$0.03} & \multicolumn{2}{c} 
{$\sim$3.5}
\end{tabular}
\end{table}

\begin{table}
\caption{Energies (relative to the three-alpha threshold) 
and full widths of low-lying natural-parity three-body 
resonances in $^{12}{\rm C}$. All quantities are given in MeV.}
\label{c12energ}
\begin{tabular}{cddddddr@{}lr@{}l}
&\multicolumn{2}{c}{MN}&\multicolumn{2}{c}{V1}&\multicolumn
 {2}{c}{V2} & \multicolumn{4}{c}{Experiment 
\protect\cite{Ajzenberg}}\\
\cline{2-3}\cline{4-5}\cline{6-7}\cline{8-11}
&E&$\Gamma$&E&$\Gamma$&E&$\Gamma$&\multicolumn{2}{c}{E}&
 \multicolumn{2}{c}{$\Gamma$} \\ 
\hline
$0^+$ & $-$10.43\tablenotemark[1] & & 
 $-$7.56\tablenotemark[1] & & $-$5.27\tablenotemark[1] & &
 \multicolumn{2}{c}{$-$7.2746\tablenotemark[1]} & & \\
& 0.64 & 0.014 & 0.71 & 0.031 & 0.83 & 0.077 & 
 0.3796&$\pm$0.0002 & (8.5&$\pm$1.0)$\times$10$^{-6}$ \\
&5.43 & 0.92 & 4.75 & 0.75 & 4.68 & 0.89 & 3.0&$\pm$0.3 & 
 3.0&$\pm$0.7 \\
&16.01 & 1.74 & 15.44 & 2.89 & 15.91 & 3.71 & 
 10.49&$\pm$0.02 & 0.08&$\pm$0.02 \\
$2^+$ & $-$7.63\tablenotemark[1] & & 
 $-$5.13\tablenotemark[1] & & $-$2.47\tablenotemark[1] & & 
 $-$2.8357&$\pm$0.0003\tablenotemark[1] & & \\
&6.39 & 1.10 & 5.55 & 1.11 & 5.49 & 1.54 & 
 3.89&$\pm$0.05 & 0.43&$\pm$0.08 \\
$3^-$ & 1.16 & 0.025 & 1.35 & 0.003 & 1.85 & 0.014 & 
 2.366&$\pm$0.005 & 0.034&$\pm$0.005 \\
&11.91 & 1.69 & 12.18 & 1.41 & 13.80 & 2.79 & 
 11.08&$\pm$0.05 & 0.22&$\pm$0.05 \\
$1^-$ & 3.71 & 0.36 & 3.72 & 0.47 & 3.82 & 0.72 & 
 3.569&$\pm$0.016 & 0.315&$\pm$0.025 \\ 
\end{tabular}
\tablenotemark[1]{Bound state.}
\end{table}


\begin{references}
\bibitem{Salpeter} E.~E. Salpeter, Phys. Rev. 88 (1952)
547; Ap. J. 115 (1952) 326; Ann. Rev. Nucl. Sci. 2 (1953)
41; Phys. Rev. 107 (1957) 516; G.~K. \"Opik, Proc. Roy.
Irish Acad. A 54 (1951) 49.
\bibitem{Hoyle} F. Hoyle, D.~N.~F. Dunbar, W.~A. Wenzel and
W. Whaling, Phys. Rev. 92 (1953) 1095; F. Hoyle, Ap. J.
Suppl. 1 (1954) 121. 
\bibitem{Ajzenberg} F. Ajzenberg-Selove, Nucl. Phys. A 490
(1988) 1. 
\bibitem{Rolfs} C.~E. Rolfs and W.~S. Rodney, Cauldrons 
in the Cosmos (The University of Chicago Press, Chicago, 
1988).
\bibitem{mac} J.~L. Visschers and R. van Wageningen, 
Phys. Lett. B 34 (1971) 455; M. Vallier\`es, H.~T. Coelho  
and T.~K. Das, Nucl. Phys. A 271 (1976) 95.
\bibitem{Marsh} S. Marsh, Nucl. Phys. A 389 (1982) 45. 
\bibitem{Fedorov} D.~V. Fedorov and A.~S. Jensen, 
Phys. Lett. B 389 (1996) 631.
\bibitem{mic} Y. Fukushima and M. Kamimura, J. Phys. 
Soc. Japan Suppl. 44 (1978) 225; M. Kamimura, Nucl. 
Phys. A 351 (1981) 456; P. Descouvemont and D. Baye, 
Phys. Rev. C 36 (1987) 54.
\bibitem{reply} A. Cs\'ot\'o, Phys. Rev. C 52 (1995) 2809. 
\bibitem{comment} A.~C. Hayes and S.~M. Sterbenz, Phys.
Rev. C 52 (1995) 2807.
\bibitem{Danilin} B.~V. Danilin and M.~V. Zhukov, 
Phys. At. Nucl. 56 (1993) 460. 
\bibitem{Matsui} Y. Matsui, Phys. Rev C 22 (1980) 2591. 
\bibitem{Eskandarian} A. Eskandarian and I.~R. Afnan, 
Phys. Rev. C 46 (1992) 2344.
\bibitem{Emelyanov} V.~G. Emelyanov, V.~I. Klimov and 
V.~N. Pomerantsev, Phys. Lett. 157B (1985) 105. 
\bibitem{CSM} Y.~K. Ho, Phys. Rep. 99 (1983) 1; N. 
Moiseyev, P.~R. Certain and F. Weinhold, Mol. Phys. 
36 (1978) 1613; Proceedings of the Sanibel Workshop Complex 
Scaling, 1978 [Int. J. Quantum Chem. 14 (1978) 343]; 
B.~R. Junker, Adv. At. Mol. Phys. 18 (1982) 207; W.~P. 
Reinhardt, Annu. Rev. Phys. Chem. 33 (1982) 223; 
Resonances--The Unifying Route Towards the Formulation of 
Dynamical Processes, Foundations and Applications in 
Nuclear, Atomic and Molecular Physics, Eds.\ E. Br\"andas 
and N. Elander, Lecture Notes in Physics vol. 325 
(Springer-Verlag, Berlin, 1989).
\bibitem{abc} J. Aguilar and J.~M. Combes, Commun. 
Math. Phys. 22 (1971) 269; E. Balslev and J.~M. 
Combes, Commun. Math. Phys. 22 (1971) 280; B. Simon, 
Commun. Math. Phys. 27 (1972) 1. 
\bibitem{Be8} A.~T. Kruppa, R.~G. Lovas and B. Gyarmati, 
Phys. Rev. C 37 (1988) 383.
\bibitem{KruppaKato} A.~T. Kruppa and K. Kat\= o, Prog. 
Theor. Phys. 84 (1990) 1145. 
\bibitem{IkedaKato} K. Kat\=o and K. Ikeda, Prog. Theor. 
Phys. 89 (1993) 623.
\bibitem{tbr} A. Cs\'ot\'o, Phys. Rev. C 49 (1994) 2244. 
\bibitem{soft} A. Cs\'ot\'o, Phys. Rev. C 49 (1994) 3035.
\bibitem{3n}A. Cs\'ot\'o, H. Oberhummer and R. Pichler, 
Phys. Rev. C 53 (1996) 1589.
\bibitem{Kato} K. Kat\= o, S. Aoyama, S. Mukai and K.
Ikeda, Nucl. Phys. A 588 (1995) 29c; S. Aoyama, S. Mukai,
K. Kat\= o and K. Ikeda, Prog. Theor. Phys. 93 (1995) 99;
Prog Theor. Phys. 94 (1995) 343. 
\bibitem{Arai} K. Arai, Y. Ogawa, Y. Suzuki and K. Varga, 
Phys. Rev. C 54 (1996) 132.
\bibitem{Afnan} I.~R. Afnan, Aust. J. Phys. 44 (1991) 201.
\bibitem{MN} D.~R. Thompson, M. LeMere and Y.~C. Tang, 
Nucl. Phys. A 268 (1977) 53; I. Reichstein and Y.~C. 
Tang, Nucl. Phys. A 158 (1970) 529. 
\bibitem{Volkov} A.~B. Volkov, Nucl. Phys. 74 (1965) 33. 
\bibitem{Kamimura} M. Kamimura, Prog. Theor. Phys. Suppl. 
62 (1977) 236. Note that this paper is unfortunately 
erroneously quoted in several of our references.  
\bibitem{pole} A. Cs\'ot\'o, R.~G. Lovas and A.~T. Kruppa, 
Phys. Rev. Lett. 70 (1993) 1389; A. Cs\'ot\'o and G.~M.
Hale, Phys. Rev. C 55 (1997) 536.
\bibitem{Li7p} A. Cs\'ot\'o and S. Karataglidis, Nucl.
Phys. A 607 (1996) 62.
\bibitem{Freer} M. Freer et.\ al., Phys. Rev. C 49 (1994) 
R1751.
\end{references}
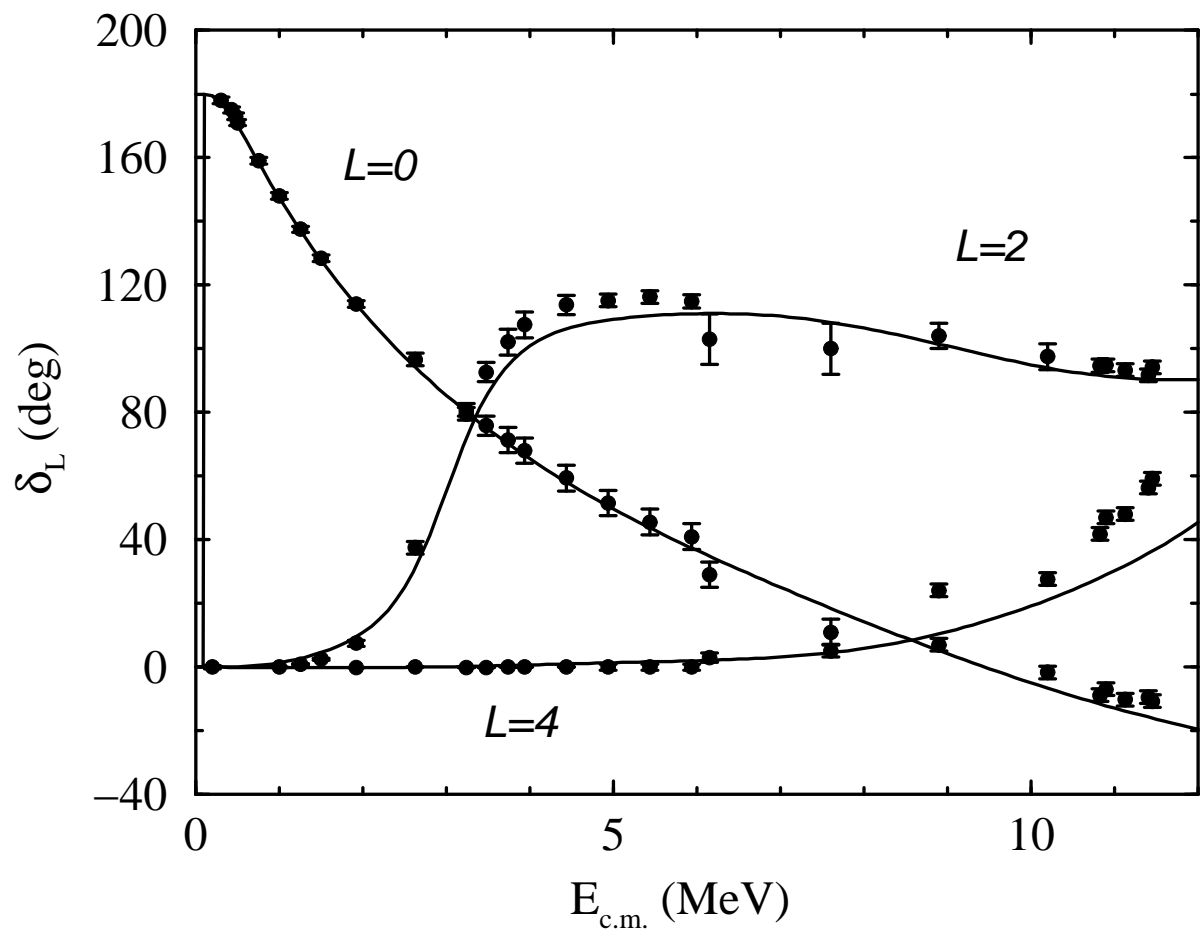
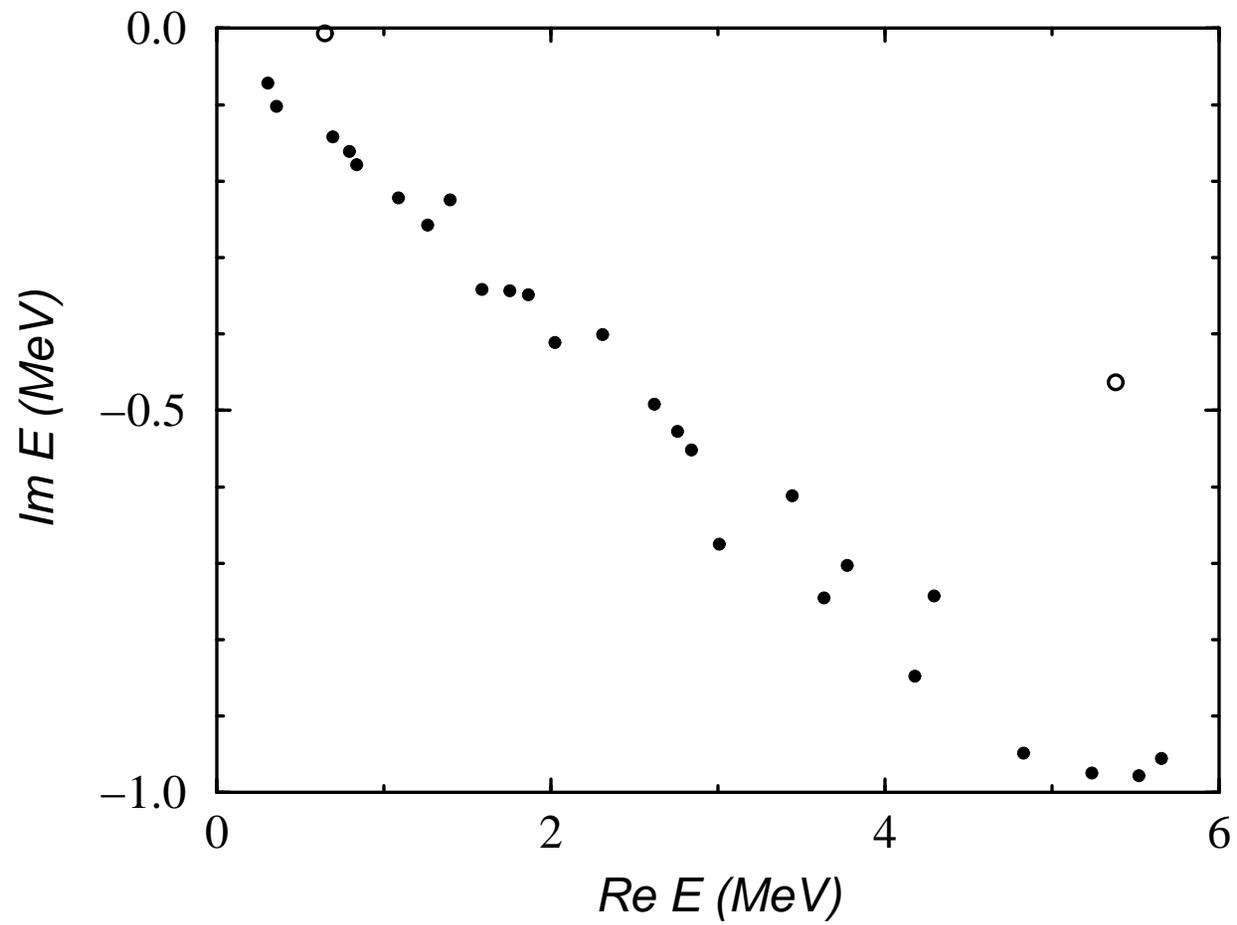
\end{document}